\documentclass[final]{eps}

\usepackage{graphicx}
\usepackage{times}
\usepackage{amsmath, amssymb}
\usepackage{ascmac}
\usepackage{natbib}

\volumenumber{00}
\publishyear{2012}
\frompage{\pageref{firstpage}}
\topage{\pageref{finalpage}}

\shorttitle{A. Sakurai \lowercase{\textit{et al.}}: SF and Extinction of
Local Galaxies as seen from AKARI and GALEX.}

\title{Star Formation and Dust Extinction Properties of Local Galaxies
as seen from AKARI and GALEX}

\author{A. Sakurai$^1$, 
T. T. Takeuchi$^1$, 
F.-T. Yuan$^1$, 
V. Buat$^2$,
and 
D. Burgarella$^2$
}
\affiliation{$^1$Division of Particle and Astrophysical Science, Nagoya University, Furo-cho, Chikusa-ku, Nagoya 464-8602, Japan\\
             $^2$Aix Marseille Universit\'{e}, CNRS, LAM (Laboratoire d'Astrophysique de
Marseille) UMR 7326, 13388, Marseille, France}
\received{xxxx xx, 2011}\revised{xxxx xx, 2012}\accepted{xxxx xx, 2012}

\abstract{
An accurate estimation of the star formation-related properties of
galaxies is crucial for understanding the evolution of galaxies.
In galaxies, ultraviolet (UV) light emitted by recently formed massive stars is
attenuated by dust, which is also produced by 
star formation (SF) activity, and is
reemitted at mid- and far- infrared (IR) wavelengths.
In this study, we investigate the star formation
rate (SFR) and dust extinction using UV and IR data. 
We selected local galaxies which are detected at AKARI
FIS $90\:\mu$m and matched the IRAS IIFSC$z$
$60\:\mu$m select catalog.
We measured FUV and NUV flux densities from GALEX images.
We examined the SF and extinction of Local galaxies 
using four bands of AKARI.
Then, we calculated FUV and total IR luminosities, and obtained the SF
luminosity, $L_{\rm SF}$, the total luminosity related to star formation activity, and the SFR. 
We find that in most galaxies, $L_{\rm SF}$ is dominated by $L_{\rm dust}$. 
We also find that galaxies with higher SF activity have a higher
fraction of their SF hidden by dust. 
In fact, the SF of galaxies with
SFRs $>20\;\rm M_{\odot}\;yr^{-1}$ is 
almost completely hidden by dust.
Our results boast a
significantly higher precision with respect to previously
published works, due to the use of much larger object samples from
the AKARI
and GALEX all sky surveys.
}


\begin{document}
\label{firstpage}
\maketitle
\copyrighttext{}

\section{Introduction}

The evolution of galaxies is one of the most fundamental problems in modern 
observational cosmology.
Since all heavy elements (elements whose atomic number is larger than
Boron) have not been produced by Big Bang Nucleosynthesis but 
 rather by stars, the investigation of star formation is related to 
the quest for the comprehending
the origin of the Earth, planets, and ourselves.

An accurate estimation of the star formation-related properties of
galaxies is crucial for the understanding of the evolution of galaxies.
The total mass of newly formed stars in a galaxy per year is referred to
as the star formation rate (SFR) 
(Takeuchi et al., 2010).
Massive stars are known to be good
indicators of star formation (SF) activity, since they have much
shorter lifetime ($\sim 10^{6-8}$~yr) than the age of galaxies and the
Universe ($\sim 10^{10}$~yr) and therefore are regarded as an ``instantaneous'' indicator of the SFR in galaxies.

Massive stars (OB stars) are hot and emit ultraviolet light.
The UV spectra is dominated by emission from massive stars.
Then, we can {\it in principle} obtain the SFR of galaxies directly by measuring 
their UV luminosity.
However, as mentioned above, stars produce heavy elements 
(or metals\footnote{$^1$ A terminology indicating elements heavier than helium.}) 
and release them through explosive phenomena
during the final phases of stellar evolution such as planetary nebulae,
supernova explosions, and some other mass ejection processes 
(e.g., Asano et al., 2012).

A significant fraction of metals are in the form of
tiny solid grain
(the typical size is smaller than $1\;\mu$m)
in the ISM, referred to as dust 
(Mathis, 1990).
This means that SF activity in galaxies is always accompanied by 
dust formation, except for the
formation of the very first stars in the Universe, 
and the dust grains are gradually mixed with the ISM.
The UV photons from massive stars are easily absorbed and/or scattered by
dust grains and re-emitted as mid- and far-infrared photons.
This is referred to as dust extinction.
We stress here that if we only measure the UV
photons from massive stars in galaxies, SF activity can be seriously 
underestimated because a significant amount of the energy is obscured by dust
(e.g., Kennicutt, 1998).
Indeed, 
Takeuchi et al. (2005a)
 have shown that a significant amount of the cosmic 
SFR density is obscured by dust and can only be observed through far-IR radiation.
They found that the fraction of hidden SF increases from 50--60\% at $z=0$ to 
$> 90$~\% at $z=1$.
This finding was confirmed and further explored by recent studies
(e.g., Murphy et al., 2011; Cucciati et al., 2012).

Many attempts have been made to explore star formation in UV and IR to
have an unbiased view of star formation in the Universe 
(e.g., Seibert et al., 2005; Martin et al., 2005;
Cortese et al., 2006; Buat et al., 2007; Lee et al., 2009; Noll et al.,
2009; Haines et al., 2011; Bothwell et al., 2011).

In order to estimate the SFR of galaxies accurately, it is ideal to combine UV-related
observables and dust-related ones.
Recently various methods on this matter have been proposed 
(e.g., Iglesias-P{\'a}ramo et al., 2006; Kennicutt et
al., 2009; Calzetti et al., 2010; Hao et al., 2011, among others).
In this study, we adopt the simplest SFR
estimator: a combination of SFRs estimated from
UV and FIR continuum radiation 
(Iglesias-P{\'a}ramo et al., 2006).
To this aim we constructed a new dataset from GALEX (UV) and AKARI (IR) observations.
Thanks to these two all-sky surveyor satellites, we have one
database which contains an unprecedented amount of 
UV-IR data of Local galaxies. 
Making use of this new database, we explore the SF- and extinction-related properties
of galaxies in the Local Universe in a similar manner to a previous work of
Takeuchi et al. (2010)
 which used the beta version of AKARI data.

This paper is organized as follows: in 
Section~2,
 we introduce the
AKARI and GALEX data and explain the construction of the infrared-selected
IR-UV dataset. In 
Section~3
 we describe the basic results of this
study, and in 
Section~4
 we interpret
the UV and IR properties
of our sample galaxies. 
Section~5
 is dedicated
 to our summary
and conclusion.

Throughout this paper we will assume $\Omega_{\rm M0} = 0.3$, $\Omega_{\Lambda0} = 0.7$ 
and $ H_0 = 70 {\rm~ km~ s^{-1}~ Mpc^{-1}}$. 
The luminosities are defined as $\nu L_{\nu}$ and expressed  in solar units  assuming  
$L_{\odot} = 3.83 \times 10^{33} {\rm ~erg~ s^{-1}}$.

\section{Data}

\subsection{AKARI}

AKARI 
was launched by JAXA (Japan Aerospace eXploration Agency) 
in February $2006$ 
(Murakami et al., 2007)\footnote{$^2$ URL: http://www.ir.isas.ac.jp/ASTRO-F/index-e.html.}.
AKARI was equipped with two imaging instruments, the
Far -infrared Surveyor 
(FIS: Kawada et al., 2007)
 and the Infrared Camera
(IRC: Onaka et al., 2007), 
together with a Fourier spectrograph 
(FTS: Kawada et al., 2008).

Before AKARI, the Infrared Astronomical Satellite, IRAS, performed the first 
all-sky survey at mid- and far-IR (MIR and FIR) wavelengths.
The IRAS all-sky survey has yielded a
vast amount of statistics for dusty galaxies in the Local Universe.
The survey provided a point source catalog (IRAS PSC)
and has long been used for extragalactic studies 
(see e.g., Soifer et al., 1987).
IRAS covered mid- and far- IR wavelength bands:
$12\;\mu$m, $25\;\mu$m, $60\;\mu$m, and $100\;\mu$m.  
Though the IRAS PSC has made a revolutionary impact on the studies of IR objects, the lack of wavebands at $\lambda > 100\;\mu$m restricted
the range of application to dusty object studies.

This is one of the reasons why AKARI was designed
to perform an IR all sky survey, especially at wavelengths longer than $100\;\mu$m.
The AKARI FIS All-Sky Survey was carried out with four photometric wavebands 
at FIR, centered at $65\;\mu$m ({\it N60}), $90\;\mu$m ({\it WIDE-S}),
$140\;\mu$m ({\it WIDE-L}), and $160\;\mu$m ({\it N160}), with a better
sensitivity and angular resolution than that of IRAS.
The point source catalogs and diffuse maps have been gradually
made public to
the astronomical community.

\subsection{GALEX}

The UV satellite GALEX (Galaxy Evolution Explorer) was launched by NASA
(the National Aeronautics and Space Administration) as one of the SMEX (Small Explorer program)
missions in April 2003\footnote{$^3$ URL: http://www.galex.caltech.edu/.}.
GALEX has two photometric bands, FUV ($1350$-$1750$~\AA, $\lambda_{\rm
mean}=1530$~\AA) and NUV ($1750$-$2750$~\AA, $\lambda_{\rm
mean}=2310$~\AA).
GALEX data products include a series of all sky surveys and deep sky
surveys in the imaging mode and partial surveys in the NUV and FUV
spectroscopic modes.
In this study we use the data of GALEX releases
GR4/GR5 all sky imaging survey (AIS) with detection limits of $19.9$ mag and $20.8$ mag in AB system
(Morrissey et al., 2007).

\subsection{Sample construction}

\subsubsection{Matching AKARI BSC with IRAS IIFSC$z$} 

As a basis for the construction of our galaxy sample we
used the AKARI FIS bright source catalog (BSC) v.$1$ from
the AKARI all sky
survey 
(Yamamura et al., 2010).
The sources in the AKARI FIS BSC v.~$1$ have a three times better S/N
ratio than the previous catalog, AKARI FIS BSC $\beta$-1 
(Yamamura et al., 2008),
which was used by 
Takeuchi et al. (2010).
This provides various advantages for the analysis of this work.
Since AKARI BSC sources contain all kinds of IR objects, like AGBs, Vega-like stars, 
H{\sc ii} regions, planetary nebulae, etc. 
(Pollo et al., 2010), the first step of this study is to construct 
a catalog of galaxies.
{}In order to have a secure sample of galaxies with redshift data, we made a cross match of AKARI 
sources with the Imperial IRAS-FSC Redshift Catalogue (IIFSC$z$), a redshift catalog recently 
published by 
Wang and Rowan-Robinson (2009).
The IIFSC$z$ is based on the IRAS Faint Source Catalog (FSC).
In the IRAS FSC, the flux density quality (FQUAL) is classified
as high ($=3$), moderate ($=2$) or upper limit ($=1$).
The IIFSC$z$ has $\mbox{FQUAL} = 3$ 
and a $\mbox{signal-to-noise ratio} > 5$ at $60\;\mu$m.
Wang and Rowan-Robinson (2009)
 selected galaxy candidates by using the following color (flux ratio) conditions:
(i) $\log (S_{100}/S_{60}) < 0.8$ (if $\mbox{FQUAL} \ge 2$ at $100\;\mu$m), 
(ii) $\log (S_{60}/S_{12}) > 0$ (if $\mbox{FQUAL} \ge 2$ at $12\;\mu$m)\footnote{Here we used
the convention of IRAS flux densities: $S_{\rm IRAS \,band} \equiv S_\nu \mbox{@IRAS band}$ [Jy].}. 
This catalog contains $60,303$ galaxies, and $90$~\% of them have spectroscopic or photometric
redshifts at $S_{60} > 0.36$~Jy.
In this sample, we put a limit on  the recession velocity of $ v > 1000\; \mbox{km}\,\mbox{s}^{-1}$, 
in order to avoid the effect of the peculiar velocities of galaxies.
Then, we matched the AKARI BSC sources with these IIFSC$z$ galaxies
using a search radius of $36$~arcsec, which corresponds to the position uncertainty
of the IRAS catalog.
If there are multiple counterpart candidates in the search radius, 
the closest one is chosen.
After concluding all of the steps described above we
obtained $6674$ IR galaxies and used them for the GALEX photometry.

\subsubsection{GALEX photometry}

We measured FUV and NUV flux densities for the $6674$ galaxies from
GALEX images by using a software package developed specially for this purpose. 
Details of this package are explained in 
Iglesias-P{\'a}ramo et al. (2006).

We performed the UV photometry as follows: 
\begin{enumerate}
 \setlength{\parskip}{0cm} 
 \setlength{\itemsep}{0cm} 
\item Cut out a $20'\times20'$ subimage from GALEX GR4/GR5 images around each
AKARI galaxy. 
\item Select the subimage with the largest exposure time. 
\item Measure the FUV and NUV flux densities from the imagelet. 
\begin{enumerate}
 \setlength{\parskip}{0cm} %
 \setlength{\itemsep}{0cm} 
\item Define the center, major axis, and minor axis to fit the ellipse
      using the NUV subimages and apply the fitted regions to
      FUV subimages.   
\item Define the sky background level around the
      major axis. 
\item Expand the ellipsoidal measured range without changing the axis ratio from the center.
\item Stop to measure when the flux density does not change more
      than the background.
\end{enumerate}
\end{enumerate}

The FUV and NUV flux densities are corrected for Galactic extinction by extracting
the  $ E(B-V)$ values from the Schlegel Map 
(Schlegel et al., 1998)
and assuming the Galactic extinction curve
calculated by 
Cardelli et al. (1989).

IR objects which did not have a counterpart in FUV
and NUV were excluded from further analysis.
The images which contain stars in the considered
region or partial galaxies (due to the GALEX image limit), were also
left out from subsequent research.
This process reduced the number of selected galaxies in
the parent sample to
$3981$.
Among them, there are galaxies which have flux
densities below the detection limits of GALEX AIS at NUV and/or FUV. 
In such a case, we replaced these unreliable values with the flux
density detection limits. 
Therefore, the effective number of galaxies which
satisfied the completeness criteria was reduced to
$3567$, and these were used
for our analysis.

\begin{figure}[t]
\centering{\includegraphics[width=0.45\textwidth]{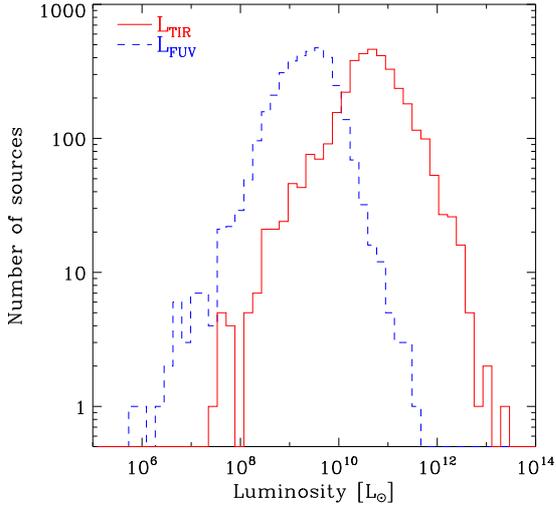}} 
\caption{$L_{\rm TIR}$ and $L_{\rm FUV}$ distributions of our sample.
Dashed and solid histograms represent the distributions of $L_{\rm FUV}$ and $L_{\rm TIR}$,
respectively.
}
\end{figure}

\section{Basic Results}

\subsection{Luminosity of galaxies}

First of all, we calculated the luminosity of sample galaxies from their measured flux density.
The flux density $S_{\nu}$ at a frequency $\nu$ and the monochromatic luminosity
$L_{\nu}$ (the luminosity per frequency) 
of an object are related through the following equation:
\begin{equation}
L_{\nu}{\rm d}\nu = 4\pi{d_{\rm L}}^{2}(z) S_{\nu}{\rm d}\nu \;,
\end{equation}
where $d_{\rm L} (z)$ is the luminosity distance of an object at redshift $z$.
Given the Hubble parameter $H(z)$ at redshift $z$
and the velocity of light $c$, $d_{\rm L}(z) $ is expressed as
\begin{equation}
d_{L}(z) = (1+z) \int_0^z \frac{c}{H( z')}{\rm d}z'
\end{equation}
where
\begin{eqnarray}
  &&H(z) = H_{0} \left[ \Omega_{\rm M 0}(1+z)^{3} \right.\nonumber \\
  &&\quad \left.-(\Omega_{\rm M 0}+\Omega_{\Lambda
			 0}-1)(1+z)^{2}+\Omega_{\Lambda 0}
			     \right]^{\frac{1}{2}} \;.
\end{eqnarray}
Then we obtain,
\begin{eqnarray}
L_{\nu_{\rm{em}}}{\rm d}\nu_{\rm{em}} &=&
 L_{(1+z)\nu_{\rm{obs}}}(1+z){\rm d}\nu_{\rm{obs}} \nonumber \\
   &=& 
 \frac{4\pi{d_{L}}^{2}S_{\nu_{\rm{obs}}}}{1+z}(1+z){\rm d}\nu_{\rm{obs}}
 \notag \\ 
&=& 4\pi{d_{L}}^{2}{\rm d}\nu_{\rm{obs}}S_{\nu_{\rm{obs}}} \;,
\end{eqnarray}
where 
\begin{equation}
\nu_{\rm{obs}} = \frac{\nu_{\rm{em}}}{1+z} \; .
\end{equation}

The AKARI- and GALEX-band luminosities are calculated with the following
formulae because of the different definitions of AKARI and GALEX photometry.
\begin{eqnarray}
&&L_{\rm AKARI \,band} \nonumber \\
&&\qquad\equiv \Delta \nu L_{\nu} \notag \\
&&\qquad= \Delta\nu\mbox{(AKARI band)} 4\pi{d_{L}}^{2}S_{\nu({\rm{@AKARI\;band}})} \;,
\end{eqnarray}
\begin{eqnarray}
&&L_{\rm GALEX\,band} \nonumber \\
&&\qquad\equiv \nu L_{\nu} \notag \\
&&\qquad= \nu\mbox{(@GALEX band)} 4\pi{d_{L}}^{2}S_{\nu{\rm{(@GALEX\;band)}}} \;.
\end{eqnarray}
Here, $\Delta \nu({\rm AKARI\;band})$ stands for the frequency range of the AKARI 
bands, and $\nu$(@GALEX band) stands for the effective frequencies of the GALEX bands.

\begin{figure}[t]
\centerline{\includegraphics[width=0.45\textwidth]{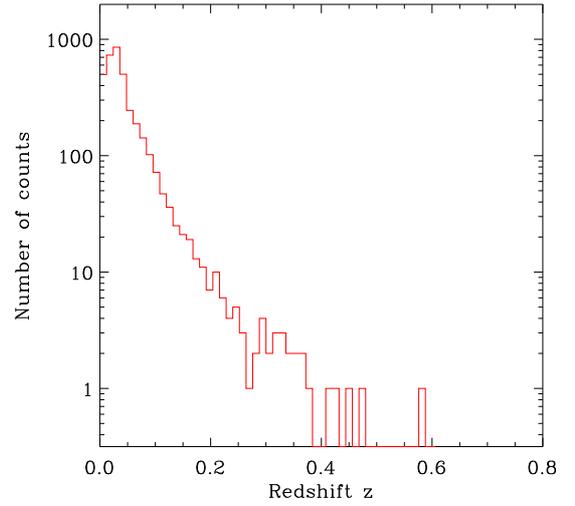}} 
\caption{Redshift distribution of the sample galaxies.
Redshift data are taken from the Imperial IRAS FSC Redshift Catalog (IIFSC$z$)
(Wang and Rowan-Robinson, 2009).}
\end{figure}
\begin{figure}[t]
\centering{\includegraphics[width=0.45\textwidth]{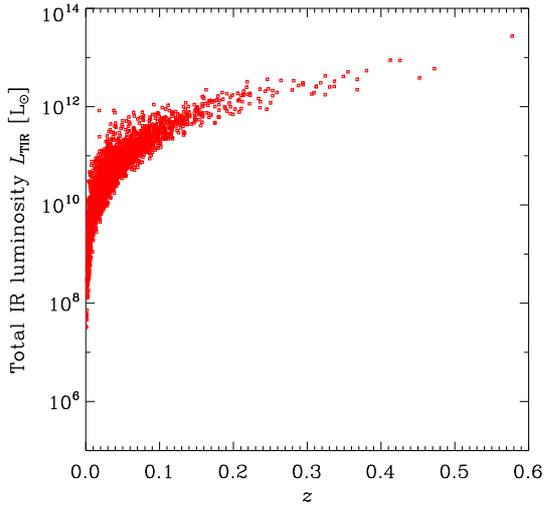}} 
\caption{The redshift-total IR (TIR) luminosity distribution.
}
\end{figure}

\subsection{Total IR luminosity}

\begin{figure}[t]
\centering{\includegraphics[width=0.45\textwidth]{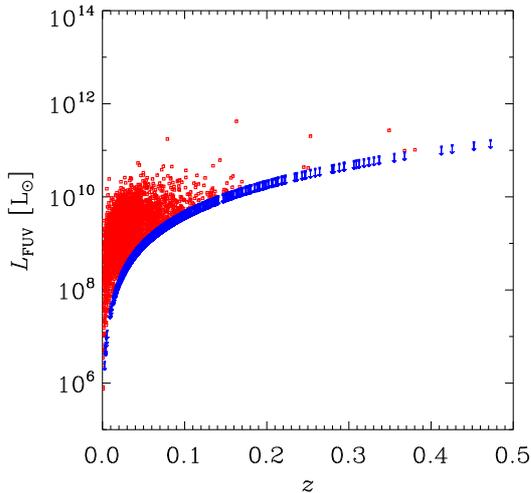}}
\caption{The redshift-FUV luminosity relations.
Open squares represent the FUV luminosities and
the downward arrows show galaxies which have FUV flux densities
below the detection limit of GALEX AIS. 
The FUV detection limit is 19.9 mag in AB system
(Morrissey et al., 2007).}
\end{figure}

We first obtained the total IR (TIR) luminosity $L_{\rm TIR}$ from AKARI
FIS bands.
Hirashita et al. (2008)
 proposed a formula to estimate the TIR luminosity
at $\lambda = 3\mbox{--}1000\;\mu$m using the flux 
densities of three AKARI bands: {\it N60}, {\it WIDE-S}, and {\it WIDE-L}, for the 
AKARI pointed observation sample of dwarf star-forming galaxies and
blue compact galaxies. 
Their formula was examined and confirmed to be valid for the AKARI FIS
All-Sky Survey galaxies by 
Takeuchi et al. (2010).
However, since the flux density at {\it N60} was noisier than
in the
wide bands, 
Takeuchi et al. (2010)
 found that an alternative formula which
makes use of only {\it WIDE-S} and {\it WIDE-L} flux densities gives 
a tighter relation than that for three bands.
Hence, in this study 
we adopted the formula of 
Takeuchi et al. (2010)
 to estimate TIR luminosity.
\begin{eqnarray}
L^{\rm{2band}}_{\rm{AKARI}} &=& \Delta\nu({\mbox{\it{WIDE-S}}})L_{\nu}(90\;\mu {\rm m}) \nonumber \\
&&\qquad+ \Delta\nu({\mbox{\it{WIDE-L}}})L_{\nu}(140\;\mu {\rm m}) \;,
\end{eqnarray}
where
\begin{equation}
\begin{split}
&\Delta\nu({\mbox{\it{WIDE-S}}}) = 1.47 \times 10^{12}\;[\rm H \rm z] \\
&\Delta\nu({\mbox{\it{WIDE-L}}}) = 0.831 \times 10^{12}\;[\rm H \rm z] \;. 
\end{split}
\end{equation}

The conversion formula from $L_{\rm AKARI}^{\rm 2band}$ to $L_{\rm TIR}$ is
\begin{equation}
\log L_{\rm{\rm TIR}} = 0.964\log L_{\rm{AKARI}}^{\rm{2band}} + 0.814
\end{equation}
The distributions of the FUV and TIR luminosities are shown in 
Fig.~1. 
We see that the two distributions are significantly different from each other.
The first reason for the difference is that our sample is AKARI
90-$\mu$m and IRAS IIFSC$z$ $60\:\mu$m selected,
i.e., they are inclined to dustier galaxies.
Then, the UV continuum of the sample galaxies is extinguished by dust, leading 
to lower luminosities in the NUV and FUV bands.
Another reason is because of the well known difference in the shape of the luminosity function
between UV and IR 
(Buat \& Burgarella, 1998; Takeuchi et al., 2005a),
 especially at the highest luminosities.
The peak luminosities at FUV and TIR are also significantly different.
More detailed analysis on the luminosity functions at UV and IR is
 presented in 
(Takeuchi et al., 2012b).

\begin{figure}[t]
\centering{\includegraphics[width=0.45\textwidth]{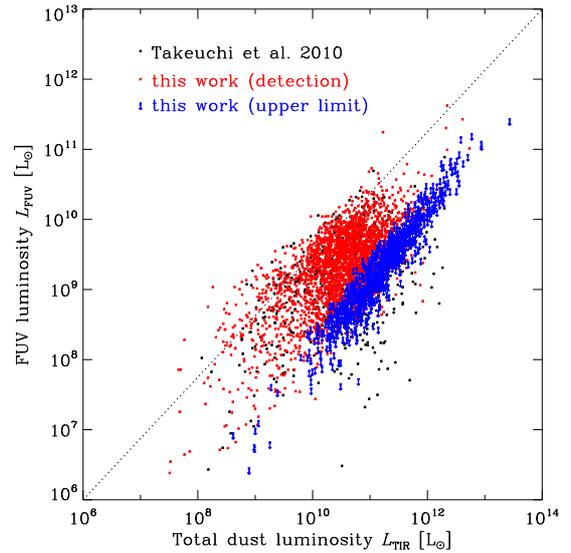}} 
\caption{Relation between $L_{\rm FUV}$ and $L_{\rm TIR}$.
The diagonal dotted line represents the case if $L_{\rm FUV}$ equals 
$L_{\rm TIR}$. Black dots represent the data from
Takeuchi et al. (2010).
Downward arrows represent galaxies which have
UV flux densities below the detection limit of GALEX.}
\end{figure} 

\begin{figure*}[t]
\centering{\includegraphics[width=0.45\textwidth]{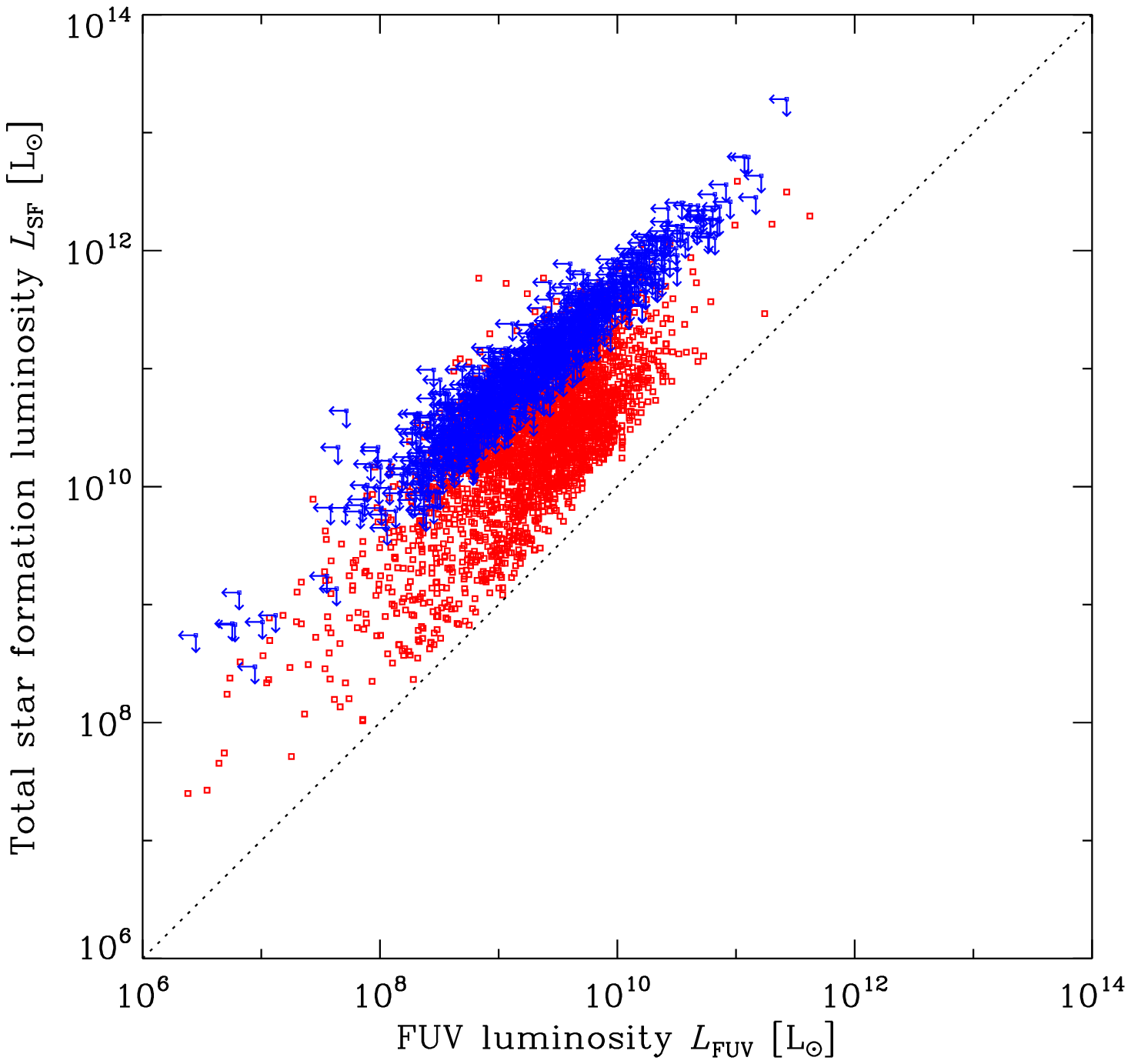}} 
{\includegraphics[width=0.45\textwidth]{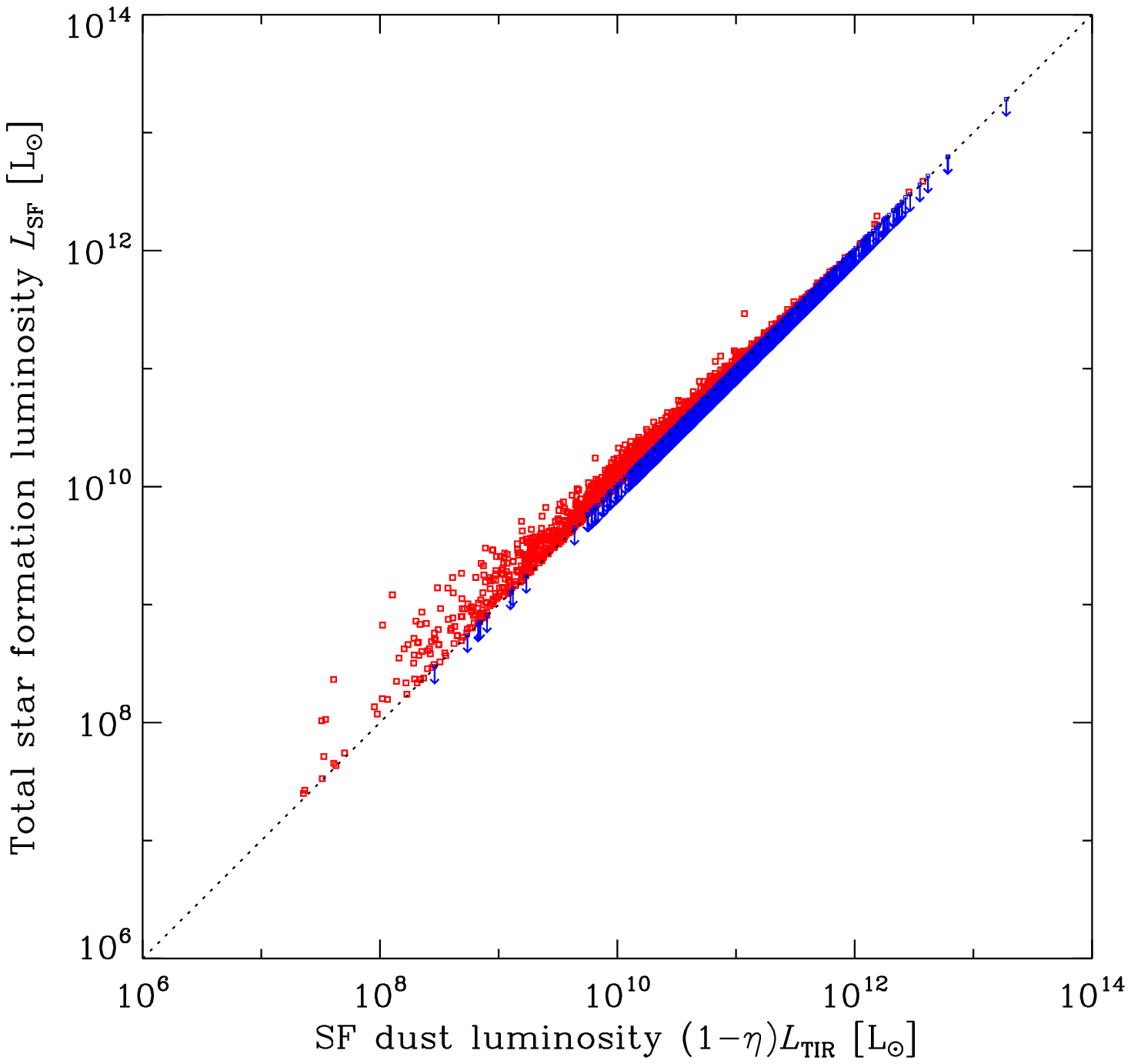}}
\caption{Contributions of FUV and SF-related dust 
luminosities to SF luminosity.
The left panel shows the contribution of FUV luminosity to 
the SF luminosity, and the right panel shows the contribution of 
the dust luminosity.
Downward and left-pointing arrows represent galaxies which have
UV flux densities below the detection limit of GALEX.}
\end{figure*}

\subsection{Redshift and luminosity distributions}

The redshift distribution of our sample is shown in 
Fig.~2.
We see that most of the sample galaxies are at low redshifts of
$z \lesssim  0.5$.
The peak of the redshift distribution is located at $z \simeq 0.02$, almost 
the same as that of the IIFSC$z$.

Figure~3
 shows the redshift-TIR luminosity distribution. 
The selection boundary due to the flux detection limit is clearly seen.
In statistical terminology this case is referred to ``truncated'' meaning
that we cannot know if there would be objects below the detection limit.

We show the relation between the redshift and FUV luminosity
in 
Fig.~4.
The small squares represent the detection at
FUV, while the downward arrows show the upper limits of the luminosities.
Due to the fact that our sample is primarily selected
at $90\;\mu$m (see 
Fig.~3)
 some galaxies detected
in IR will be invisible at UV bands (see 
Fig.~4).
This case is referred to as ``censored''.
This difference is important when we try to estimate a luminosity function.
We discuss this issue elsewhere 
(Takeuchi et al., 2012b).

\subsection{Sample completeness}

Now we examine the completeness of the final sample.
{From} number counts of the sample, we expect that the flux limit is
$S_{90} = 0.45$~Jy.
The redshift completeness of the sample is tested by
 $V/V_{\rm max}$ statistics (Schmidt 1968; Rowan-Robinson 1968). 
Here $V$ is the volume enclosed in a sphere whose radius is the distance
of a certain source in the sample, and $V_{\rm max}$ is the volume enclosed
in a sphere whose radius is the largest distance at which the
source can be detected. 
If the sample is complete, $V/V_{\rm max}$ of the sample galaxies is expected to 
be distributed uniformly between 0 and 1, with an average $\langle V/V_{\rm max} \rangle = 0.5$ 
and a standard deviation $(12)^{-1/2}$. 
For our final sample, we obtain $\langle V/V_{\rm max} \rangle = 0.51 \pm 0.27$, i.e., 
the sample can be regarded as complete down to 0.45~Jy. 
Thus, even if we have set additional conditions to have redshifts from IIFSC$z$, our sample
is complete above this flux limit. 
The following analysis are all based on this limit.

\subsection{Star formation luminosity}

Now we compare FUV luminosity (mainly from massive stars) with TIR luminosity 
(mainly from dust) for the sample galaxies.
Figure~5
 shows that our sample galaxies are much more
luminous at FIR than at FUV.
Downward arrows in 
Fig.~5
 represent galaxies which have
UV flux densities below the detection limit of GALEX. 

Here, we define the star formation luminosity, $L_{\rm SF}$, as 
the total luminosity contributed only by massive stars.
The SF luminosity is expressed as
\begin{equation}
L_{\rm{SF}} \equiv L_{\rm{\rm FUV}} + (1-\eta)L_{\rm{\rm TIR}} \;,
\end{equation}
where $\eta$ is the fraction of IR emission produced by dust heated
by old stars, which is not related to the current SF.
We adopted $30~\%$ for this fraction 
(Hirashita et al., 2003)
 in the Local
Universe. 
Buat et al. (2011)
 have shown that $\eta$ is almost constant for a wide range of
$L_{\rm TIR}$ for galaxies.
Their result supports the assumption of a constant $\eta$.

Figure~6
 shows the contributions of FUV and SF-related dust 
luminosities to the SF luminosity.
Here, the SF dust luminosity stands for $(1-\eta)L_{\rm TIR}$.
The left panel in 
Fig.~6
 shows the contribution of FUV luminosity to 
the SF luminosity, and the right panel shows the contribution of 
the SF dust luminosity.
It is clearly seen that the galaxies are significantly above the diagonal line, and  
the scatter is large in the left panel.
This comes directly from the fact that $L_{\rm TIR}$ is much larger than
$L_{\rm FUV}$.
In contrast, the right panel of 
Fig.~6
 shows a very tight correlation
between $(1 - \eta) L_{\rm TIR}$ and $L_{\rm SF}$. 
We should note that, by definition,  $(1 - \eta) L_{\rm TIR} < L_{\rm SF}$. 
However, as seen in 
Fig.~6,
 the galaxies with upper limits exactly
delineate the diagonal line.
For these galaxies, the SF activity is almost completely hidden by dust.
We discuss this hidden SF in more detail in the next section.

\section{Discussion}

\subsection{Star formation rate}

Since we measure the emission from massive stars, we need to
convert the number of massive stars to the total number of stars.
We use an initial mass function (IMF) for this conversion.
The IMF represents the number of newly formed stars per mass.
We assume the Salpeter IMF 
(Salpeter, 1955).

With the spectral evolutionary synthesis model Starburst99 
(Leitherer et al., 1999), and 
assuming a constant SFR over $10^8$~yr, solar
metallicity, and the Salpeter IMF (mass range $0.1M_\odot \mbox{--}100 M_\odot$), we obtain the relation between the SFR and $L_{\rm FUV}$ as
(Iglesias-P{\'a}ramo et al., 2006)
\begin{equation}
\log {\rm{SFR_{\rm FUV}}} = \log L_{\rm FUV} - 9.51 \; .
\end{equation}
The relation between SFR and TIR luminosity is 
\begin{equation}
\log {\rm{SFR_{dust}}} = \log L_{\rm TIR} - 9.75 + \log (1-\eta) \; ,
\end{equation}
where we have assumed that all stellar UV light is absorbed by dust.
And we obtain the formula under the same assumption for both the
star formation history (SFH) and the IMF as those of the FUV.
We obtain the following formula to calculate the total SFR:
\begin{equation}
{\rm{SFR}} = {\rm SFR_{\rm FUV}} + {\rm{SFR_{dust}}} \;,
\end{equation}
where ${\rm SFR_{\rm FUV}}$ is the SFR estimated from directly observable UV luminosity,
and ${\rm{SFR_{dust}}}$ is that estimated from the dust luminosity.

The fraction of $\rm SFR_{\rm FUV}$ to the total SFR for the sample galaxies
is shown in 
Fig.~7.
The scatter of the fraction is very large at $\rm SFR<20\;M_{\odot}yr^{-1}$. 
However, there is a sudden drop at $\rm SFR>20\;M_{\odot}yr^{-1}$. 
This means that the fraction of the hidden SF strongly depends on the SFR,
and galaxies with higher SFRs are more strongly extinguished by dust.

We also find some outliers which have high SFRs and high $\rm SFR_{FUV}$
fractions. 
A possible explanation is that they might harbor quasars/AGNs. 
Since the UV energy of these objects is generated from accretion disks
around central black holes, the efficiency of the energy release is much 
higher than for usual SF.
As such, quasars tend to have a pointlike strong UV source in the center.
In this case, $L_{\rm UV}$ is not related to the SF activity
in a simple way so
we do not discuss this further in this work. 
We cross matched the galaxies to the quasars/AGNs of 
V{\'e}ron-Cetty \& V{\'e}ron (2010)
 and the identified galaxies are represented by the black
crosses on the red or blue symbols in 
Fig.~7. 
We find that 246 objects in our sample are known
quasars/AGNs. The ratio of quasars/AGNs to galaxies with SFR $\rm \leq, 20 M_{\odot}yr^{-1}$
is $5.9 \%$ and the ratio of quasars/AGNs to galaxies with SFR $\rm > 20
M_{\odot}yr^{-1}$ is $11 \%$.
We also find that $6/9$ galaxies which have $\rm SFR > 20 M_{\odot}yr^{-1}$
and $\rm SFR_{FUV}/SFR > 0.3$ are quasars/AGNs.
Thus, we can safely conclude that normal star-forming galaxies
(without contamination of UV flux from AGNs) have
small $\rm SFR_{FUV}/SFR$ ratios.
This is also consistent with the result by Totani et al.\ (2011) from
a different aspect.

Figure~8
 shows the ratio $\rm SFR_{FUV}/SFR_{dust}$ as
a function of $L_{\rm SF}$.
The similar ratio using $\rm SFR_{NUV}$ was presented in 
Iglesias-P{\'a}ramo et al. (2006).
We can relatively safely compare their result with ours, because $\rm SFR_{NUV}$
and $\rm SFR_{FUV}$ are calibrated from the same model, and the total IR luminosity 
in 
Iglesias-P{\'a}ramo et al. (2006)
 is estimated with a formula proposed by 
Dale et al. (2001),
which was proved to be a close approximation to our $L_{\rm TIR}$ 
(see Takeuchi et al., 2005b).
We observe a larger scatter in 
Fig.~8 than 
Iglesias-P{\'a}ramo et al. (2006).
This may be because our sample is much larger and contains more extremely
dusty objects.

Figure~9
 shows the relation between SFR and $L_{\rm
TIR}$.
We find a very tight correlation.
Comparing our result with the result of 
Bothwell et al. (2011),
 there are more 
galaxies in the regime with higher $L_{\rm TIR}$ ($L_{\rm TIR} \gtrsim
10^{12}\;{\rm L}_{\odot}$) and higher SFR (SFR $\gtrsim 10^{2}\;{\rm
M}_{\odot}{\rm yr}^{-1}$), and few galaxies in the
lower $L_{\rm TIR}$ ($L_{\rm TIR} \lesssim 10^{8}\;{\rm L}_{\odot}$)
 and lower SFR (SFR $\lesssim 10^{-1}\;{\rm M}_{\odot}{\rm yr}^{-1}$) regime.
This is because 
Bothwell et al. (2011)
added low-luminosity local volume galaxies to
their IR- and UV-selected samples, and they made a volume correction
on these data, while our sample is only IR-selected.

We also see that galaxies with FUV flux densities below the {\sl GALEX} 
detection limit delineate the same trend which was reported by 
Bothwell et al. (2011).
Further, galaxies with ${\rm SFR} < 10^{-1} \; M_\odot\,\mbox{yr}^{-1}$ start to
deviate from the general trend, also consistent with 
Bothwell et al. (2011).
This confirms that galaxies with low SFR have larger
contributions from $\mbox{SFR}_{\rm FUV}$.

\begin{figure}[t]
\centering{\includegraphics[width=0.45\textwidth]{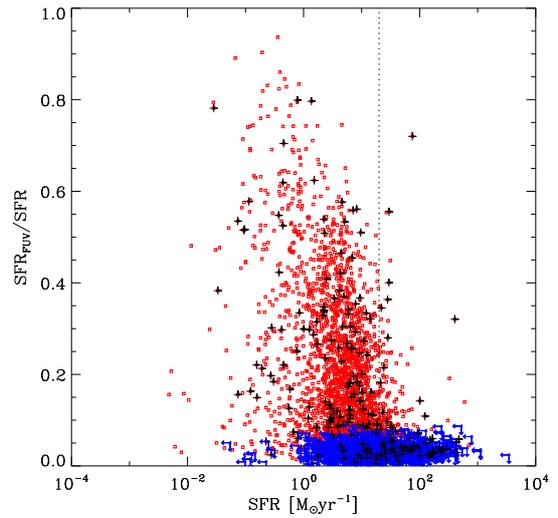}} 
\caption{Contribution of the fraction of $\rm SFR_{FUV}$ to the
 total SFR. The red and blue symbols are the same as in
Fig.~6. 
The black crosses represent the quasars and active galaxies cross
 matched to the objects by 
V{\'e}ron-Cetty \& V{\'e}ron (2010).
The dotted line represents the total SFR where $\rm SFR_{FUV}/SFR$
 drops suddenly for increasing SFR.}
\end{figure}
\begin{figure}[t]
\centering{\includegraphics[width=0.45\textwidth]{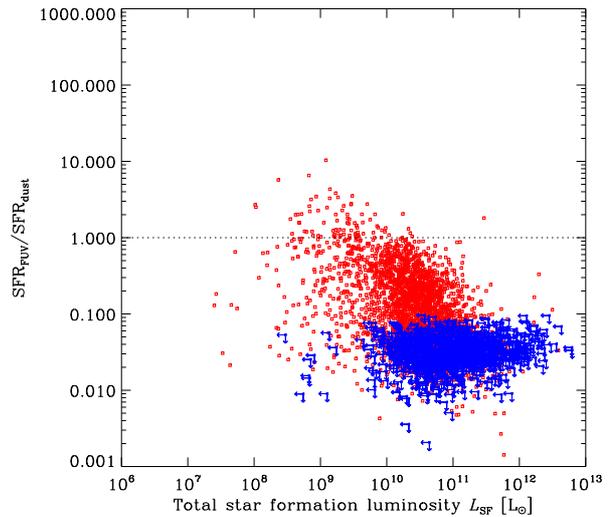}} 
\caption{Ratio between $\rm SFR_{FUV}$ and $\rm SFR_{TIR}$ as a
 function of
star-formation luminosity $L_{\rm SF}$. The red and blue symbols are the same as in
Fig.~6.} 
\end{figure}
\begin{figure}[t]
\centering{\includegraphics[width=0.45\textwidth]{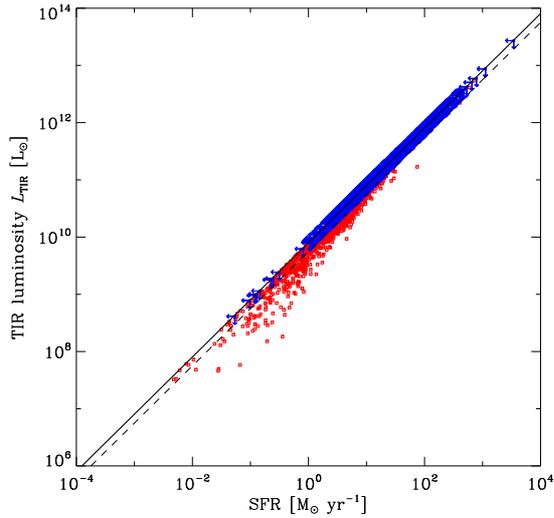}}
\caption{Relation between SFR and $L_{\rm TIR}$.
The red and blue symbols are the same as in
Fig.~6.
The solid line represents the $L_{\rm TIR}$-SFR scaling
 given by 
Eq.~(13). The dotted line represents the $L_{\rm FIR}$-SFR
 scaling given by 
Iglesias-P{\'a}ramo et al. (2006).}
\end{figure}
\begin{figure}[t]
\centering{\includegraphics[width=0.45\textwidth]{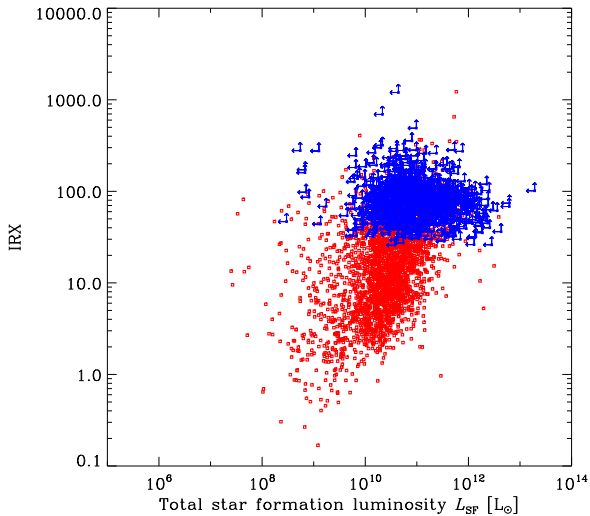}} 
\caption{Relation between $L_{\rm SF}$ and the infrared excess
 $\mbox{IRX} \equiv L_{\rm TIR}/L_{\rm FUV}$.
Upward and left-pointing arrows 
 represent galaxies which have UV flux densities below the detection limit of GALEX. 
The scatter of the IRX is large.}
\end{figure}
\begin{figure}[t]
\centering{\includegraphics[width=0.45\textwidth]{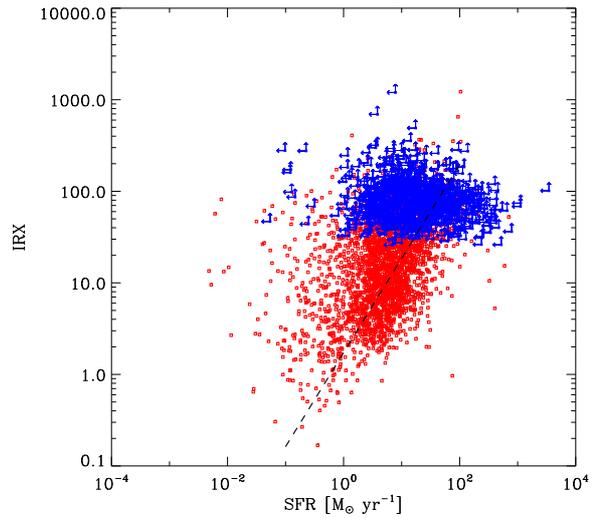}}
\caption{Relation between SFR and IRX.
The red and blue symbols are the same as in
Fig.~10. 
The dotted line represents the IRX-SFR relation
 derived for star forming $z \sim 0$ galaxies by 
Overzier et al. (2011).}
\end{figure}

\subsection{Dust extinction}
The relation between $L_{\rm SF}$ and the infrared excess $\mbox{IRX} \equiv L_{\rm TIR}/L_{\rm FUV}$, 
known as an indicator of extinction as a function of IR luminosities,
 is shown in 
Fig.~10.
Figure~10 has the trend that galaxies which have larger
$L_{\rm SF}$ have high IRX, consistent
with Fig.~7 of 
Buat et al. (2007).

Rather naturally, we expected strong dust extinction for the current sample 
because they are IR-selected.
We find a similar result to that of 
Buat et al. (2005)
and 
Takeuchi et al. (2010).

Figure~11
 shows the relation between SFR and IRX.
Again, we compare it with Fig.~10 of 
Bothwell et al. (2011). 
Since we plot the sample galaxies directly without
a volume correction, 
it is not straightforward to compare the distribution of galaxies on this figure.
Even so, we can see that we have less galaxies
in the low SFR (SFR$<
10^{-1}\;{\rm M_{\odot}yr^{-1}}$) and small attenuation (IRX$<1$) regime
compared with 
Bothwell et al. (2011).
As for 
Fig.~9, this comes from
the effects of sample selection and
our sample is inclined to more luminous galaxies because of the IR-selection.
The global trend is, however, almost the same for 
Bothwell et al. (2011).

Figure~12
 shows the relation between
the FUV$-$NUV color and IRX.
The galaxies which have UV flux densities below the detection limit are
not plotted.
We show the dependence on the two different IR luminosities, $L_{\rm IRAS 60\mu m}$ and
$L_{\rm TIR}$, for comparison with various previous studies.
In general, $L_{\rm TIR}$ is smaller than $L_{\rm IRAS 60\mu m}$ as you
can also see in 
Fig.~9.
We can compare the left panel of 
Fig.~12
 with Fig.~15 of 
Takeuchi et al. (2010)
 directly, and 
find a similar trend.
By contrast, $L_{\rm TIR}$ is more directly connected to the definition of IR luminous
galaxies, LIRGs ($10^{11}{\rm L_{\odot}} \leq L_{\rm TIR} < 10^{12} {\rm
L_{\odot}}$) and ultra IR luminous galaxies, ULIRGs ($10^{12} {\rm
L_{\odot}} \leq L_{\rm TIR}$).
Indeed, more galaxies in the right panel are identified as the IR luminous galaxies
according to the criterion of $L_{\rm TIR}$ dependence. 
Solid curves in each panel represents the revised IRX-$\beta$ relation obtained from 
the GALEX-AKARI measurement of the same UV-luminous starbursts as 
Meurer et al. (1999), 
proposed by 
Takeuchi et al. (2012a). 
Most IR luminous galaxies are above the curve in
each panel.
This trend was discovered by 
Goldader et al. (2002)
 and followed by subsequent studies 
(e.g., Buat et al., 2005; Takeuchi et al., 2010, among others).
On the other hand, non IR luminous galaxies, i.e. lower extinction
galaxies,  follow this curve. 
This may be explained as follows: 
Takeuchi et al. (2012a)
 used the same original galaxy sample of 
Meurer et al. (1999), which
was selected for central UV-luminous intense starbursts. 
Since these galaxies tend to have lower dust attenuation than LIRGs/ULIRGs, they are
similar to the lower-IR luminosity galaxies in our sample having significant UV fluxes, as we 
have seen in 
Fig.~9.  
Hence, we conclude that low-luminosity IR galaxies have a common attenuation strength
with UV-luminous starbursts.

\begin{figure*}[t]
\centering{\includegraphics[width=0.45\textwidth]{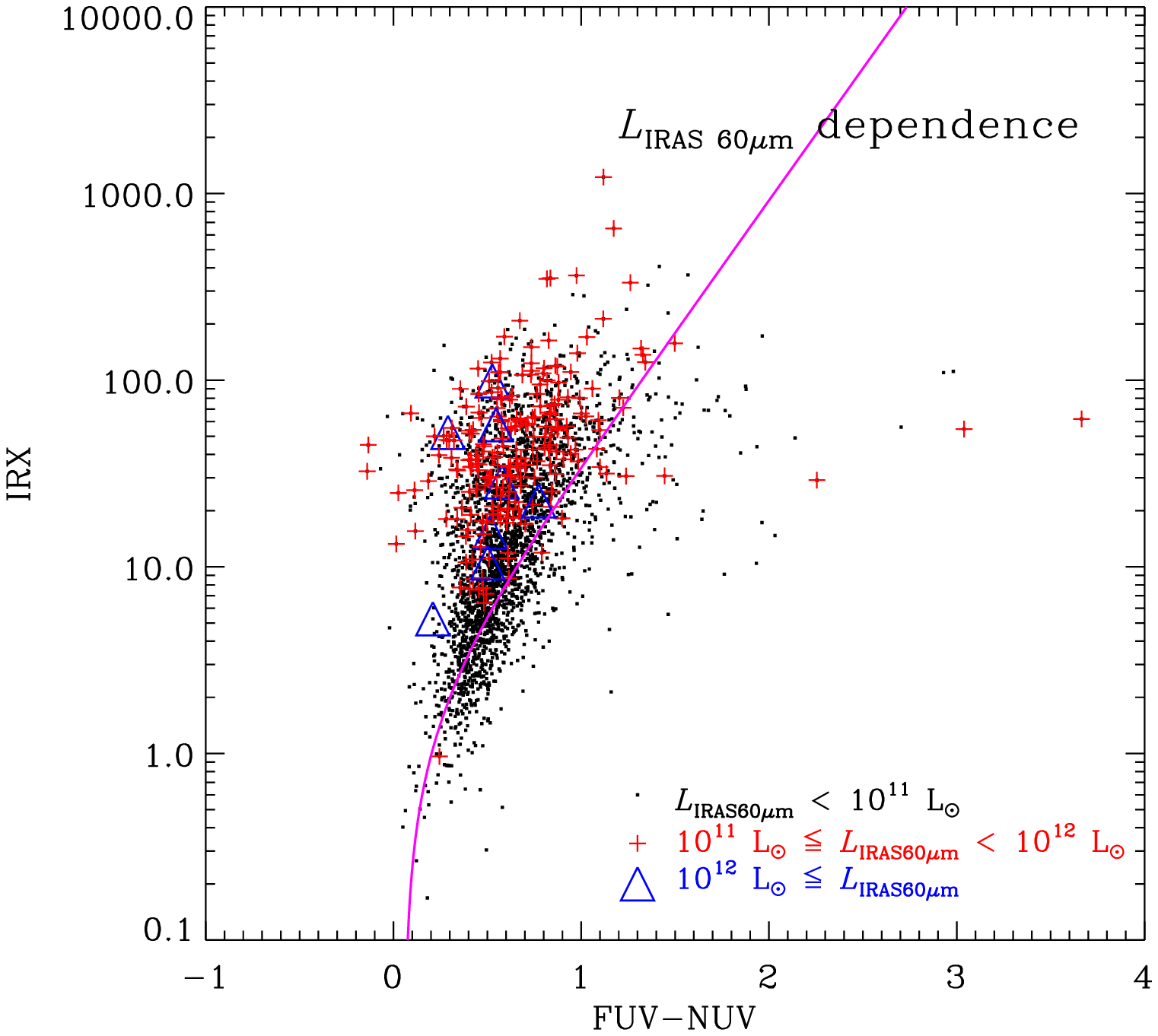}} 
{\includegraphics[width=0.45\textwidth]{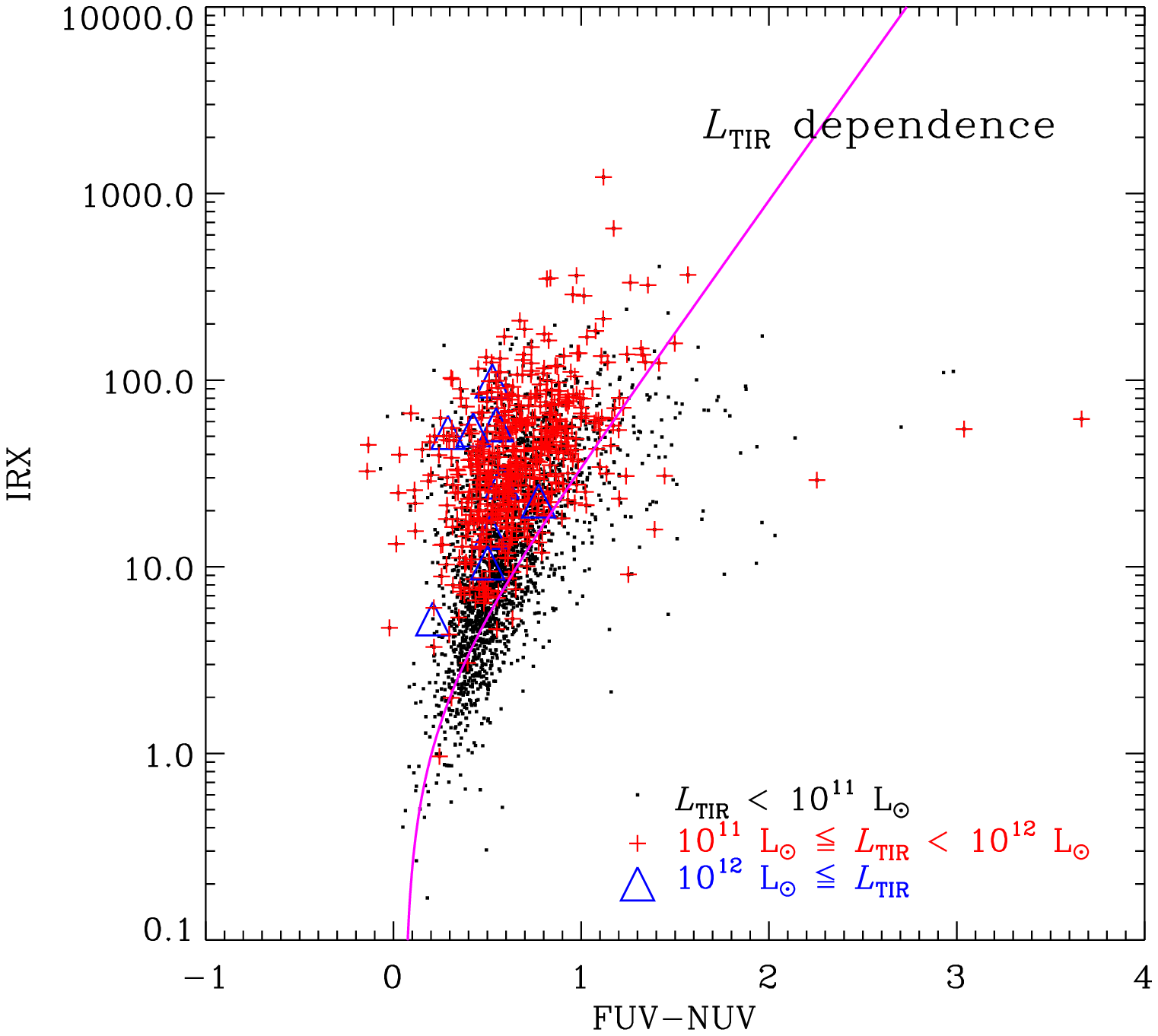}}
\caption{Relation between FUV$-$NUV color and IRX.
The left panel shows the relation as a function of IRAS $60\mu m$ luminosity,
 $L_{\rm IRAS\, 60\,\mu{\rm m}}$. 
The right panel is the same but with total IR luminosity, $L_{\rm TIR}$. 
Dots: galaxies with $L_{\rm
 IR} < 10^{11}\;{\rm L}_{\odot}$, 
crosses: galaxies with
 $10^{11}\;{\rm L}_{\odot} \leq L_{\rm IR} < 10^{12}\;{\rm L}_{\odot}$, 
triangles: galaxies with $L_{\rm IR} \leq 10^{12}\;{\rm L}_{\odot}$. 
Here $L_{\rm IR}$ stands for $L_{\rm IRAS\,60\,\mu{\rm m}}$ in the left panel and
$L_{\rm TIR}$ in the right panel.
The solid curve in each panel represents the IRX--UV-color relation of 
UV-luminous starbursts 
(Takeuchi et al., 2012a).}
\end{figure*}

\section{Conclusions}

We analyzed star formation-related properties of Local galaxies using AKARI and GALEX data.
The summary and conclusions of this study are as follows:
\begin{enumerate}
 \setlength{\parskip}{0cm} 
 \setlength{\itemsep}{0cm} 
\item The star formation luminosity, $L_{\rm SF}$, is dominated by the 
emission from dust related to SF activity, $(1 - \eta)L_{\rm TIR}$.
\item The contribution of ultraviolet luminosity, $L_{\rm FUV}$, has a
larger scatter than that of the contribution of $(1 - \eta)L_{\rm TIR}$.
\item It is difficult to estimate the star formation activity only from
the relation between $L_{\rm SF}$ and $L_{\rm FUV}$ because of the small
contribution of $L_{\rm FUV}$.
\item Galaxies with higher SF activity ($\rm \mbox{SFR} > 20 \;M_{\odot}\;\mbox{yr}^{-1}$) have
a higher fraction of their SF hidden by dust.
\item We examined the relation between IRX, 
$L_{\rm TIR}/L_{\rm FUV}$, and FUV$-$NUV color. 
Among the current sample, low-IR luminosity galaxies ($< 10^{11}\;L_\odot$) follow 
the relation for UV-luminous starbursts proposed by 
Takeuchi et al. (2012a).

\end{enumerate}
These conclusions are consistent with those of 
Takeuchi et al. (2010).
However, we find that the dispersion  
in various relations they obtained suffered from the noise of the AKARI BSC $\beta$-1 catalog.
In this study, the S/N is three times better so we do not share the same
problem.
Thus, we can safely conclude that the above properties are general
features of Local star-forming dusty galaxies.

\acknowledgments
This work is based on observations with AKARI, a JAXA project with the participation of ESA. 
TTT has been supported by Program for Improvement of Research
Environment for Young Researchers from Special Coordination Funds for
Promoting Science and Technology, and the Grant-in-Aid for the Scientific
Research Fund (20740105, 23340046, 24111707) commissioned by the MEXT. 
AS, TTT, and FTY have been partially supported from the Grand-in-Aid for the Global
COE Program ``Quest for Fundamental Principles in the Universe: from
Particles to the Solar System and the Cosmos'' from the Ministry of
Education, Culture, Sports, Science and Technology (MEXT) of Japan.
We are grateful to Stefan Noll and the anonymous referee for their
helpful comments which improved the presentation and content of
this paper.
We deeply thank Noriko Tsuchiya for her early contribution to this
work, and Jennifer M.\ Stone and Aleksandra Solarz for their help
in the improvement of English.


\email{A. Sakurai (e-mail: sakurai.akane@e.mbox.nagoya-u.ac.jp)}
\label{finalpage}
\lastpagesettings
\end{document}